\documentclass[11pt]{article}
\usepackage{graphicx}
\begin{document}

\begin{center}
{\LARGE \bf Rotations associated with Lorentz boosts} \\

\vspace{7mm}

S. Ba{\c s}kal \footnote{electronic
address:baskal@newton.physics.metu.edu.tr} \\
Department of Physics, Middle East Technical University,\\
06531 Ankara, Turkey
\vspace{5mm}

Y. S. Kim\footnote{electronic address: yskim@physics.umd.edu}\\
Department of Physics, University of Maryland,\\
College Park, Maryland 20742, U.S.A.\\

\end{center}

\begin{abstract}
It is possible to associate two angles with two successive non-collinear 
Lorentz boosts.  If one boost is applied after the initial boost, 
the result is the final boost preceded by a rotation called the 
Wigner rotation.  The other rotation is associated with Wigner's 
$O(3)$-like little group.  These two angles are shown to be different.  
However, it is shown that the sum of these two rotation angles is equal 
to the angle between the initial and final boosts.  This relation is 
studied for both low-speed and high-speed limits. 
Furthermore, it is noted that the two-by-two matrices which are under 
the responsibility of other branches of physics can be interpreted 
in terms of the transformations of the Lorentz group, or vice versa.  
Classical ray optics is mentioned as a case in point.
\end{abstract}

\newpage

\section{Introduction}\label{intro}

The Wigner rotation is known as a by-product of two successive
Lorentz boosts in special relativity. The earliest manifestation of the
Wigner rotation is the Thomas precession which we observe in atomic spectra.  
Thomas formulated this problem thirteen years before the appearance of 
Wigner's 1939 paper~\cite{thomas26,wig39}. The Thomas effect in nuclear 
spectroscopy is mentioned in Jackson's book on electrodynamics~\cite{jack99}.
Indeed, the Wigner rotation is the key issue in many branches of physics
involving Lorentz boosts~\cite{yama81}.

Recently, the Lorentz group has become an important scientific language
in both quantum and classical optics.  The theory of squeezed states is
a representation of the Lorentz group~\cite{yuen76,knp91}.  Optical
instruments are unavoidable in modern physics, and they are based on
classical ray optics.  It is gratifying to observe that the Lorentz
group, through its two-by-two representation, is the basic underlying
scientific language for ray optics, including polarization
optics~\cite{hkn97}, interferometers~\cite{hkn00}, lens
optics~\cite{sudar85,bk03}, laser cavities~\cite{bk02}, and multi-layer
optics~\cite{monz96}. 

It is possible to perform mathematical operations of the Lorentz group 
by arranging optical instruments.  For instance, the group
contraction is one of the most sophisticated operations in the Lorentz
group, but it has been shown recently that this can be achieved through
focal process in one-lens optics~\cite{bk03}.  
Since there are many mathematical operations in optical sciences 
corresponding to Lorentz boosts, the Wigner  rotation becomes one of 
the important issues in classical and quantum optics.

If we perform two Lorentz boosts in different directions, the result
is not a boost, but is a boost preceded or followed by a rotation.
This rotation is commonly known as the Wigner rotation.
However, if we trace the origin of this word, Wigner introduced the
rotation subgroup of the Lorentz group whose transformations leave the
four-momentum of a given particle invariant in its rest frame.  The
rotation can however change the direction of its spin.  Indeed, Wigner
introduced the concept of ``little group'' to deal with this type of
problem.  Wigner's little group is the maximum subgroup of the Lorentz
group whose transformations leave the four-momentum of the particle
invariant.  If the particle is moving, we can go to the Lorentz frame
where it is at rest, perform a rotation without changing the momentum,
and then come back to the original Lorentz frame.  These transformations
leave the momentum invariant.  We shall hereafter call this little-group
rotation ``WLG rotation.''

The question then is whether the Wigner rotation, as understood in
the literature, is the same as the WLG rotation.
This question was raised by Han {\it et al.} in their paper on
Thomas precession and gauge transformations, but they have not made
any attempt to clarify this issue~\cite{hks87cqg}.  The present authors 
raised this question again in their paper on laser cavities~\cite{bk02}.  
They first noted that the two-by-two matrix formulation of lens optics is
a representation of the Lorentz group, and then showed that the light
beam performs one little-group rotation as it goes through one cycle
in the cavity.  Then they showed that the Wigner rotation and the
WLG rotation are different, but those rotation angles were related for
the special case of the Thomas precession.

The purpose of this paper is to establish the same relation for the most
general case.  We establish the difference between those two angles, and
then show that they satisfy a complementary relation. 
In spite of the simplicity in concept, the calculations of these angles
are not trivial. 

Every relativistic problem has two important limits.  One
is the non-relativistic limit, and the other is the light-like limit where
the momentum of the particles becomes infinitely large.  We also study
these angles and their relation for the two limiting cases.

We note that the $SL(2,C)$, the group of unimodular 
two-by-two matrices, is the universal covering group of the Lorentz group, 
having the same algebraic property as the four-by-four representation 
of the Lorentz group.  
Although, for completeness we have included the expressions of the four-by-four 
transformation matrices, needless to say, their two-by-two counterparts
can be expressed in a much more compact way. 
Furthermore, and more important than that, within the SL(2,C) formalism 
these matrix calculations can be applied to the two-by-two beam transfer 
matrices and the two-by-two lens matrices in classical ray optics.  Indeed, 
our basic motivation for the present paper came from our experience in ray 
optics. Thus, the group  $SL(2,C)$ provides not only a topological base for 
the Lorentz group, but also concrete calculational tools for various branches 
of physics.

In Sec.~\ref{formul}, we consider two different rotations associated with
two successive non-collinear Lorentz boosts.
One is the Wigner rotation, and the other is the rotation associated with
Wigner's little group.  It is shown that the addition of these two angles
is equal to the angle between the direction of the first boost and the
the final boost.
In Sec.~\ref{compu}, using the two-by-two matrices, we explicitly calculate 
those angles in terms of the parameters of the initial Lorentz boosts.  
In Sec.~\ref{example}, we give some illustrative examples to show the
dependence of the angles on the boost parameters.
In Sec.~\ref{two}, we explain how special relativity and ray optics 
find a common mathematical ground through their two-by-two matrix 
formalism.

\section{Two Different Angles}\label{formul}

\begin{figure}[thb]
\centerline{\includegraphics[scale=0.7]{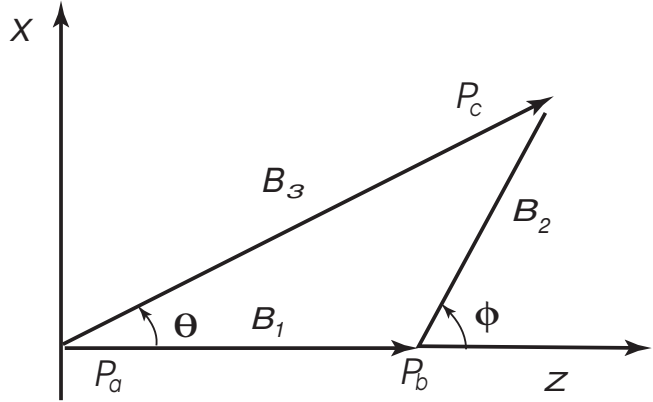}}
\vspace{6mm}
\caption{Two successive Lorentz boosts.  Let us start from a particle
at rest.  If we make  boost $B_{1}$ along the $z$ direction and another
$B_{2}$ along the direction with makes an angle of $\phi$ with the $z$
direction, the net result is not $B_{3}$, but $B_{3}$ preceded by a
rotation.  This rotation is known as the Wigner rotation.}\label{angle11}
\end{figure}

In the literature, the Wigner rotation comes from two successive Lorentz
boosts performed in different directions.  If we boost along the $z$ axis
first and then make another boost along the direction which makes an angle
$\phi$ with the $z$ axis on the $zx$ plane as shown in Fig.~\ref{angle11},
the result is another Lorentz boost preceded by a rotation.
This rotation is known as the Wigner rotation in the literature.

In the metric $(t,z, x, y)$, the rotation matrix which performs a
rotation around the $y$ axis by angle $\phi$ is
\begin{equation}\label{r44}
  R(\phi) = \pmatrix{1 & 0 & 0 & 0 \cr
         0 & \cos\phi & -\sin\phi & 0 \cr
         0 & \sin\phi & \cos\phi & 0 \cr 0 & 0 & 0 & 1} ,
\end{equation}
and its inverse is $R(-\phi)$.

The boost matrix requires two parameters.  One is the boost parameter,
and the other is the angle specifying the direction. We shall use the
notation
\begin{equation}
B(\phi, \eta)
\end{equation}
as the matrix performing a boost along the direction which makes an
angle of $\phi$ with the boost parameter $\eta$.  The boost matrix along
the $z$ direction takes the form
\begin{equation}\label{b44}
B(0, \eta) = \pmatrix{\cosh\eta & \sinh\eta & 0 & 0 \cr
           \sinh\eta & \cosh\eta & 0 & 0 \cr
           0 & 0 & 1 & 0 \cr  0 & 0 & 0 & 1} .
\end{equation}
If this boost is made along the $\phi$ direction, the matrix is
\begin{equation}\label{rb44}
B(\phi, \eta) = R(\phi)~ B(0, \eta)~ R(-\phi),
\end{equation}
and its inverse is $B(\phi, -\eta)$.


Let us start with a massive particle at rest whose four-momentum is
\begin{equation}\label{4a}
P_{a} = (m, 0, 0, 0),
\end{equation}
where $m$ is the particle mass.  If we apply the boost matrix $B(0, \eta)$
to the four-momentum it becomes
\begin{equation}\label{4b}
P_{b} = m (\cosh\eta, \sinh\eta, 0, 0) .
\end{equation}
If we apply another boost $B(\phi, \lambda)$, the four-momentum 
takes the form
\begin{equation}\label{4c}
P_{c} = m \left(\cosh\xi, (\sinh\xi)\cos\theta,
           (\sinh\xi) \sin\theta, 0 \right) .
\end{equation}
The kinematics of these transformations is illustrated in
Fig.~\ref{angle11}.
Then, we can consider the successive boosts
\begin{equation}
B(\theta, -\xi)~B(\phi, \lambda)~B(0, \eta) .
\end{equation}
If this matrix is applied to $P_{a}$ of Eq.(\ref{4a}), it brings
back to $P_{a}$.  This means that the net effect is a rotation
$R(\omega)$, which does not change the four-momentum of the particle
in its rest-frame.
This aspect is commonly written in the literature as
\begin{equation}\label{wrot11}
B(\phi,\lambda) B(0, \eta) = B(\theta, \xi) R(\omega) ,
\end{equation}
where the matrices $ B(0, \eta), B(\phi, \lambda)$ and
$B(\theta, \xi)$ correspond to
$B_{1}, B_{2}$ and $B_{3}$ in Fig.~\ref{angle11} respectively.

The product of the two boost matrices appears to be one boost matrix
on the right-hand side in Fig.~\ref{angle11}, but there must be a
rotation matrix $R(\omega)$ to complete the mathematical identity.
This rotation is known as the Wigner rotation in the literature.
\begin{equation}\label{wrot22}
R(\omega) = B(\theta, -\xi)~B(\phi, \lambda)~ B(0, \eta) .
\end{equation}

\begin{figure}[thb]
\centerline{\includegraphics[scale=0.7]{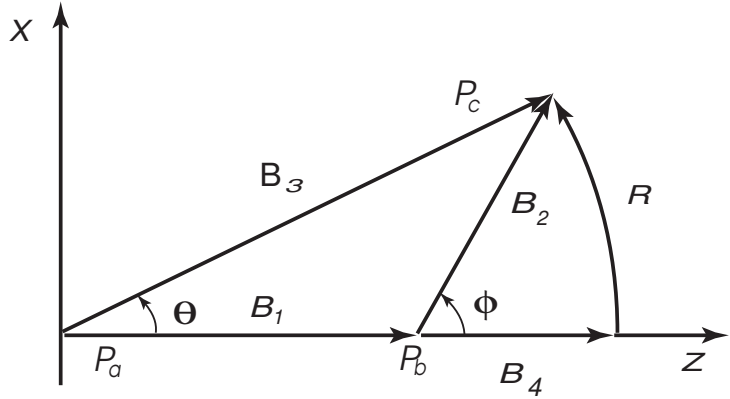}}
\caption{Closed Lorentz boosts. Initially, a massive particle is at
rest with its four momentum $P_{a}$.  The first boost $B_{1}$ brings
$P_{a}$ to $P_{b}$.  The second boost $B_{2}$ transforms $P_{b}$ to
$P_{c}$.  The third boost $B_{3}$ brings $P_{c}$ back to $P_{a}$.
The particle is again at rest. The net effect is a rotation
around the axis perpendicular to the plane containing these three
transformations.  We may assume for convenience that $P_{b}$ is
along the $z$ axis, and $P_{c}$ in the $zx$ plane. The rotation
is then made around the $y$ axis.} \label{angle22}
\end{figure}

Let us consider a different transformation to obtain $P_{c}$ from $P_{b}$.
We can first boost the system by $B(0, \xi - \eta)$, and rotate 
it by $R(\theta)$.  The boost along the same direction does not change
the helicity of the particle.  The rotation $R(\theta)$ is also
a helicity preserving transformation.  This route is illustrated in
Fig.~\ref{angle22}.  Helicity-conserving transformations has been
discussed extensively in the literature~\cite{bk02,hks85}.

\begin{figure}[thb]
\centerline{\includegraphics[scale=0.7]{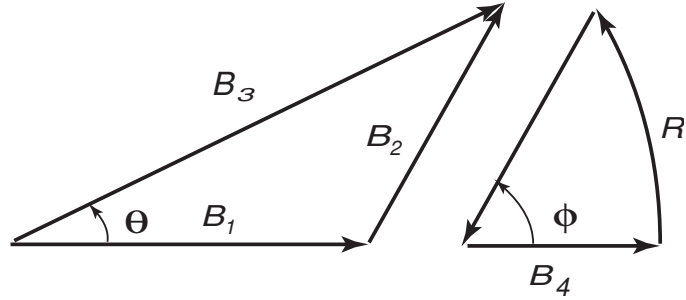}}
\caption{Addition of the angles.  This figure consists of the
Fig.\ref{angle11} and the kinematics corresponding to the  $D$ matrix
of Eq.(\ref{d11}).  This figure also illustrates the addition rule of
Eq.(\ref{addition}).} \label{angle33}
\end{figure}

There are now two different ways of obtaining $P_{c}$ from
$P_{b}$.  If we choose the second route, and come back
using $B(\phi, -\lambda)$, the net effect is
\begin{equation}\label{d11}
D(\eta,\lambda,\phi) = B(\phi, -\lambda) [R(\theta) B(0, \xi - \eta)] .
\end{equation}
This transformation leaves the four-momentum $P_{b}$ given
in Eq.(\ref{4b}) invariant.  This ``loop'' transformation is illustrated 
in Fig.(\ref{angle33}).

This is not the only way to leave the given the four-momentum unchanged.
If we apply the boost $B(0, -\eta)$ to the four-momentum $P_{b}$ of
Eq.(\ref{4b}), the result would be the four-momentum $P_{a}$ of
Eq.(\ref{4a}).  This is the four-momentum of the particle at rest.  This
four-momentum is invariant under three-dimensional rotations.  This
is precisely what Wigner observed in defining the $O(3)$-like
rotation group for massive particles~\cite{wig39}.  After performing
a rotation which leaves $P_{a}$ invariant, we can boost the momentum
back to $P_{b}$ by applying $B(0, \eta)$.  The net effect is
\begin{equation}\label{d22}
B(0, \eta) R(\alpha) B(0, -\eta).
\end{equation}
This is the original definition of Wigner's little group which leaves
$P_{b}$ invariant.  The rotation matrix $R(\alpha)$ represents a
three-dimensional rotation matrix.

We now demand that the little group transformation of Eq.(\ref{d22})
is the same as the $D$ matrix of Eq.(\ref{d11}).  Then,
\begin{equation}\label{alpha2}
B(0, \eta) R(\alpha) B(0, -\eta) = B(\phi, -\lambda)
~R(\theta) B(0, \xi - \eta) .
\end{equation}
This determines the angle $\alpha$ as:
\begin{equation}\label{wlgr}
R(\alpha) = B(0, -\eta) B(\phi, -\lambda)~R(\theta) B(0, \xi) .
\end{equation}
This is the WLG rotation angle as defined in Sec.~\ref{intro}.

Let us next consider the product $R(\omega)~R(\alpha)$, where
$R(\omega)$ and $R(\alpha)$ are from Eq.(\ref{wrot22}) and
Eq.(\ref{wlgr}), respectively.  Then
\begin{equation}
R(\omega)~R(\alpha) = R(\theta) ,
\end{equation}
which leads to
\begin{equation}\label{addition}
\alpha + \omega = \theta .
\end{equation}

It is interesting to note that the above relation does not depend
on the direction of the $B(\theta,\xi)$, nor does on the boost 
parameters $\eta$ and $\lambda$.

The purpose of this paper is to study consequences of the
above relation.

\section{Computation of the rotation angles}\label{compu}

In this section, we compute both Wigner rotation and WLG rotation
angles.  The two-by-two representation of the rotation matrix
corresponding to the four-by-four expression of Eq.(\ref{r44}) is 
\begin{equation}\label{r22}
R(\phi) = \pmatrix{\cos(\phi/2)  & - \sin(\phi/2) \cr
     \sin(\phi/2)  &  \cos(\phi/2) },
\end{equation}
while the boost matrix given in Eq.(\ref{b44}) becomes
\begin{equation}\label{b22}
B(0, \eta) = \pmatrix{e^{\eta/2} & 0  \cr 0 & e^{-\eta/2} }.
\end{equation}
Let us use $B(\phi, \eta)$ for the boost along the direction
which makes an angle $\phi$ with the $z$ axis.  Then it takes the form
\begin{equation}\label{rb22}
\pmatrix{\cosh(\eta/2) + (\cos\phi)\sinh(\eta/2) &
  (\sin\phi)\sinh(\eta/2) \cr
  (\sin\phi)\sinh(\eta/2) &
   \cosh(\eta/2) - (\cos\phi)\sinh(\eta/2) } .
\end{equation}
Uing these two-by-two expressions, we can complete all the computations
for the transformation matrices given in Sec.~\ref{formul}.

Let us go to the calculation of the Wigner rotation angle defined in
Eq.(\ref{wrot11}).  We can compute $\xi, \theta$ and $\omega$ in terms of  
$\eta, \lambda$ and $\phi$, by requiring that the right-hand side of
Eq.(\ref{wrot22}) be a rotation matrix~\cite{hkn88,hhk89}.  
The result of this calculation is
\begin{eqnarray}\label{tan11}
&{}& \cosh\xi = \cosh\eta~\cosh\lambda
            + \sinh\eta~\sinh\lambda~\cos\phi ,    \nonumber \\[2ex]
&{}&  \tan\theta = {\sin\phi [\sinh\lambda +
          \tanh\eta (\cosh\lambda -1) \cos\phi] \over
        \sinh\lambda \cos\phi + \tanh\eta [1 +
        (\cosh\lambda - 1)\cos^{2}\phi]} ,\nonumber \\[2ex]
&{}&  \tan\omega =  {2 (\sin\phi) [\sinh\lambda\sinh\eta  +
  C_{-} \cos\phi] \over
    C_{+} + C_{-}\cos(2\phi) +
     2 \sinh\lambda\sinh\eta\cos\phi } ,
\end{eqnarray}
with
\begin{equation}
C_{\pm} = (\cosh\lambda \pm 1) (\cosh\eta \pm 1) .
\end{equation}

As for the angle $\alpha$, we first compute the boost 
parameter $\beta$ of $B_{4}$ in terms of $\eta, \lambda$ and $\phi$ 
as 
\begin{equation}
\tanh\beta=\frac
{f-\tanh\eta \,(1+\tanh\eta\tanh\lambda\cos\phi)}
{(1+\tanh\eta\tanh\lambda\cos\phi)-f\,\tanh\eta} ,
\end{equation}
and then obtain the $D$ 
matrix of Eq.(\ref{d11}) which takes the form
\begin{equation}\label{loopm11}
 D(\eta,\lambda,\phi) = \pmatrix{[(f + g)/2f]^{1/2} &
  [h_{+} (f - g)/ 2f]^{1/2} \cr
  [h_{-}(f - g)/2f]^{1/2}  & [(f + g)/2f]^{1/2}   }  ,
\end{equation}
where
\begin{eqnarray}
&{}& f = \frac{\sqrt{(\cosh\eta \cosh\lambda  +
\sinh\eta\sinh\lambda\cos\phi)^{2} - 1}}
{\cosh\eta\cosh\lambda} , \nonumber \\[2ex]
&{}& g = \tanh\eta +\tanh \lambda \cos\phi , \nonumber \\[2ex]
&{}& h_{\pm} = { 1 \pm \tanh\eta \over 1 \mp \tanh\eta} .
\end{eqnarray}
The four-by-four counterpart of $ D(\eta,\lambda,\phi)$ 
is of the form
\begin{equation} \label{loopm44}
\pmatrix{ 
[f\,\cosh^{2}\eta-g\,\sinh^{2}\eta]/f & n\,(g-f)/f 
         & -\kappa/f & 0  \cr
    -n\,(g-f)/f   & [-f\,\sinh^{2}\eta + g\,\cosh^{2}\eta]/f
       &  -s/f & 0   \cr
   -\kappa/f & s/f
    &g/f  & 0  \cr 
0 & 0 & 0 & 1},
\end{equation}
where
\begin{equation}
\kappa=\tanh\eta\tanh\lambda\sin\phi,\qquad n=\sinh\eta \cosh\eta,
\qquad s=\tanh\lambda\sin\phi.
\end{equation}
On the other hand, the left-hand side of Eq.(\ref{d11}) is
$B(0,\eta) R(\alpha) B(0, -\eta)$, which takes the form
\begin{equation}\label{loopm22}
\pmatrix{\cos(\alpha/2) & -e^{\eta/2}\sin(\alpha/2) \cr
 e^{-\eta/2}\sin(\alpha/2) & \cos(\alpha/2)} .
\end{equation}
Now, in view of Eq.(\ref{alpha2}), we can calculate the angle $\alpha$ by 
equating Eq.(\ref{loopm11}) and Eq.(\ref{loopm22}).  The result is
\begin{equation}\label{tan22}
\tan\alpha=\frac{\tanh\lambda \sin\phi}
     {\sinh\eta + \cosh\eta \tanh\lambda \cos\phi}.
\end{equation}

We can check the addition law given in Eq.(\ref{addition})
by computing
\begin{equation}
\tan(\alpha + \omega) = { \tan\alpha + \tan\omega \over
1 - (\tan\alpha) \tan\omega }.
\end{equation}
After completion of this calculation using $\tan\omega$ and
$\tan\alpha$ of Eq.(\ref{tan11}) and Eq.(\ref{tan22}) respectively,
we end up with $\tan\theta$ of Eq.(\ref{tan11}).

In terms of the velocity of the particle, $\tanh\eta = v/c$.  This means
that $v = c\eta$ in the slow-speed limit.  If the particle speed
approaches the speed of light, $\tanh\eta$ becomes 1.
Let us consider the velocity additions in both cases.  If $\eta$ and
$\lambda$ are both small, the expressions in Eq.(\ref{tan11}) become
\begin{eqnarray}\label{tan33}
&{}& \xi^{2} = \eta^{2} + \lambda^{2}
            + \eta\lambda~\cos\phi ,    \nonumber \\[2ex]
&{}&  \tan\theta = {\lambda \sin\phi \over
       \eta + \lambda \cos\phi } ,
       \nonumber \\[2ex]
&{}&  \alpha = \theta,  \nonumber \\[2ex]
&{}&  \omega =  0 .
\end{eqnarray}
These expressions are consistent with the addition rules of
non-relativistic kinematics.  The Wigner rotation does not exist because
$\omega =  0.$

If $\eta$ and $\lambda$ are small, the system becomes the
non-relativistic case.  If $\eta$ becomes infinitely large,
we are dealing with light-like particles.  In the limit of large $\eta$
we have:
\begin{eqnarray}\label{tan44}
&{}& \xi = \eta + \ln(\cosh\lambda +
                     \sinh\lambda~\cos\phi) ,   \nonumber \\[2ex]
&{}&  \tan\theta = {\sin\phi [\sinh\lambda +
          (\cosh\lambda -1) \cos\phi] \over
          \sinh\lambda \cos\phi +
          [1 + (\cosh\lambda - 1)\cos^{2}\phi]} ,\nonumber \\[2ex]
&{}&  \alpha = 0 ,    \nonumber \\[2ex]
&{}&  \omega = \theta .
\end{eqnarray}

As for the $D$ matrix of Eq.(\ref{loopm11}), it becomes
\begin{equation}\label{rot11}
\pmatrix{\cos(\alpha/2) & -\sin(\alpha/2) \cr
\sin(\alpha/2) & \cos(\alpha/2)},
\end{equation}
in the limit of small $\eta$ and $\lambda$.  This matrix represents a
rotation by an angle $\alpha$ around the $y$ axis.  This form is
consistent with the expressions given in Eq.(\ref{loopm22}).

Let us go back to the original definition of Wigner's little group for
massive particles.  For a given massive particle, moving along the
$z$ direction, we can bring the particle to its rest frame.  Then we
can perform a rotation without changing the four-momentum of the particle.
However, the direction of its spin changes.  We can bring back the
particle to its original momentum by applying a boost matrix.
This is what is happening in Eq.(\ref{d22}).  If the amount of boost is
very small, the little-group transformation is a rotation as given in
Eq.(\ref{rot11}).

For massless particles, it is not possible to bring the particle to its
rest frame.  The best we can do is to align the $z$ axis along the
direction of the momentum.  In his original paper~\cite{wig39}, Wigner
observed that the subgroup of the Lorentz group which dictates the
internal space-time symmetry is locally isomorphic to the
two-dimensional Euclidean group, with one rotational and two
translational degrees of freedom.  The rotational degree of freedom
corresponds to the helicity, but the translation-like degrees were
left unexplained.

Let us look at the $D$ matrix of Eq.(\ref{loopm11}).  When $\eta$
becomes very large, and $\tanh\eta$ approaches  1, this matrix becomes
\begin{equation}\label{loopm33}
  D(\lambda,\phi) = \pmatrix{1 & u \cr 0 & 1} ,
\end{equation}
where
\begin{equation}
  u = \frac{2 \tanh \lambda \sin \phi}{1+\tanh \lambda \cos \phi}.
\end{equation}
Similarly, when $\tanh\eta$ approaches to 1,
the $D$ matrix of Eq.(\ref{loopm44}) becomes:
\begin{equation}
D = \pmatrix{ 1 + u^2/4 & -u^2/2 & -u & 0  \cr
           u^2/2  & 1 - u^2/2 &  -u & 0   \cr
           -u & u & 1 & 0  \cr 0 & 0 & 0 & 1} .
\end{equation}
This expression was given in Wigner's original paper~\cite{wig39}, and
corresponds to one of the translation-like transformations for the
massless particle, but its physical interpretation as a gauge
transformation was first given by Janner and Janssen~\cite{janner71}.
Indeed, this matrix had a stormy history~\cite{wein64,hk81ajp,knp86},
and its full story had not been told until 1990 when Kim and Wigner
presented a cylindrical picture of the $E(2)$-like little
group for massless particles~\cite{kiwi90jm}.  This little group
as a generator of gauge transformations is also an interesting subject
in general relativity~\cite{banerj01}.

Furthermore, it is interesting to see that the expression of the $D$
matrix can be obtained as a large-$\eta$ limit of the Lorentz-boosted
rotation of Eq.(\ref{loopm22}).  This is a procedure known as the
group contraction which In{\"o}n{\"u} and Wigner introduced to physics in
1953~\cite{inonu53}.  In their paper, In{\"o}n{\"u} and Wigner considered
a two-dimensional plane tangent to a sphere, and observed that a small
area on the spherical surface can be regarded as a two-dimensional
plane with the two-dimensional Euclidean symmetry.  Indeed, the
In{\"o}n{\"u}-Wigner contraction is the contraction of the rotation group
$O(3)$ to the two-dimensional Euclidean group.

Since the symmetry groups for massive and massless particles are
locally isomorphic to the rotation and Euclidean groups respectively,
it was expected that the symmetry group of massless particle could
be obtained through a contraction procedure.  This aspect also has a
history~\cite{bacry68,hks83pl}, but the problem had not been completely
clarified when Kim and Wigner in 1990 introduced a cylindrical
symmetry for massless particles~\cite{kiwi90jm}.  The question was
that there are two-translational degrees of freedom while there is only
one gauge degree of freedom.

\section{Illustrative Examples}\label{example}

The calculations of Sec.~\ref{compu} become simpler if the angle $\phi$
takes a special value.  If this angle is such that the boost parameter
$\xi$ remains the same as $\eta$, this transformation is responsible for 
Thomas precession.  For this simpler case, the addition law 
$\theta = \alpha + \omega $ was noted in our earlier paper~\cite{bk02}.  
The formulas of Eq.(\ref{tan11})
and Eq.(\ref{tan22}) become
\begin{eqnarray}
&{}& \cosh\xi = \cosh\eta ,    \nonumber \\[2ex]
&{}&  \tan\theta = \tan \theta  ,\nonumber \\[2ex]
&{}& \tan\alpha = \frac{2\sin\theta \cosh\eta}
   {\sinh^2\eta + (1 + \cosh^2\eta)\cos\theta)}, \nonumber \\[2ex]
&{}&  \tan\omega = 
\frac{\sin\theta[\cos\theta (\cosh\eta - 1)^{2}+\sinh^{2}\eta}
   {\cos\theta[\cos\theta [(\cosh\eta - 1)^{2}+\sinh^{2}\eta]+ 2\cosh\eta} .
\end{eqnarray}
In our earlier paper~\cite{bk02}, we calculated $\alpha$  and $\omega$ in 
terms $\theta$ and $\eta$, instead of $\lambda$ and $\eta$.

\begin{figure}[thb]
\centerline{\includegraphics[scale=0.9]{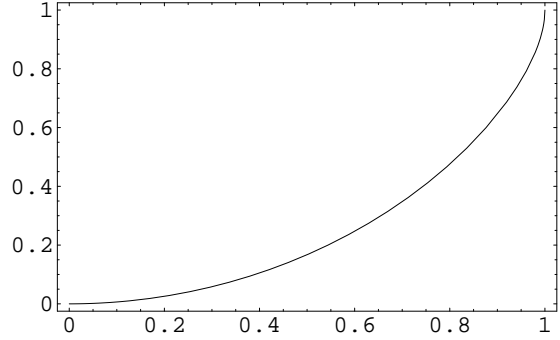}}
\caption{The ratio of the angle $\omega$ to the angle $\theta$ and
a function of $\tanh\eta$, which becomes  one as $\eta$ becomes very
large.  The ratio is zero at $\eta = 0$, while it becomes one as
$\eta$ approaches infinity. This was expected from the limiting
cases discussed at the end of Sec~\ref{compu}.} \label{angle44}
\end{figure}

If the angle $\phi$ is $90^{o}$, the expressions of Eq.(\ref{tan11}) and
Eq.(\ref{tan22}) also become simpler, and the kinematics becomes quite
transparent~\cite{hks87jm}.  The angles are

\begin{eqnarray}\label{tan55}
&{}& \cosh\xi = \cosh\eta~\cosh\lambda,    \nonumber \\[2ex]
&{}&  \tan\theta = {\sinh\lambda \over \tanh\eta} ,\nonumber \\[2ex]
&{}&\tan\alpha={\tanh\lambda \over \sinh\eta }, \nonumber \\[2ex]
&{}&  \tan\omega = { \sinh\lambda\sinh\eta
 \over  \cosh\eta  + \cosh\lambda } .
\end{eqnarray}

We can now plot the above expressions as $\eta$ goes from zero to
infinity, or as $\tanh\eta$ goes from zero to 1, for a given value of
$\lambda$.  Let us try the case with $\lambda = \eta$.  Then the
expressions become
\begin{eqnarray}\label{tan66}
&{}& \cosh\xi = \cosh^{2}\eta,    \nonumber \\[2ex]
&{}&  \tan\theta =  \cosh\eta ,  \nonumber \\[2ex]
&{}&\tan\alpha={1 \over \cosh\eta }, \nonumber \\[2ex]
&{}&  \tan\omega = {\sinh\eta~\tanh\eta \over 2 } .
\end{eqnarray}
In terms of $\tanh\eta$,
\begin{eqnarray}\label{tan77}
&{}& \cosh\xi = {1 \over 1 - \tanh^{2}\eta},    \nonumber \\[2ex]
&{}&  \tan\theta = {1 \over \sqrt{1 - \tanh^{2}\eta}} ,\nonumber \\[2ex]
&{}&\tan\alpha=\sqrt{1 - \tanh^{2}\eta }, \nonumber \\[2ex]
&{}&  \tan\omega = {\tanh^{2}\eta \over 2 \sqrt{1 - \tanh^{2}\eta}} .
\end{eqnarray}
If we plot the angle $\theta$ against $\tanh\eta$, it starts with
$45^{o}$ at $\eta = 0$.  The angle monotonically increases to $90^{o}$ as
$\tanh\eta$ reaches $1$.  We can also plot $\alpha$ and $\omega$ to
appreciate the addition rule given in Eq.(\ref{addition}).


\section{Physics of two-by-two matrices}\label{two}
According to Eugene Wigner, quantum mechanics is the physics of Fourier
transformations, and special relativity is the physics of Lorentz
transformations.

In our recent papers, we formulated classical ray optics in terms of the
two-by-two matrix representation of the Lorentz group, meaning that special 
relativity and ray optics has found a common  mathematical formulation.  
It was noted that optical instruments can serve as analogue computers 
for special relativity through the use of those two-by-two matrices.  
Most of the calculations done in this present paper, particularly the 
group contraction mentioned in Sec.~\ref{compu}, can be carried out by 
optical instruments~\cite{bk03}.  Indeed, the motivation of this work is 
substantially based on the results of the papers written earlier by the 
present authors on ray optics.

Coherent and squeezed states in quantum optics can be formulated in 
terms of Wigner functions defined in two-dimensional phase space and 
linear canonical transformations~\cite{knp91}.  
Many physical theories are formulated as two-level problems.   Most of the
soluble models in physics take the form of coupled harmonic oscillators.
Needless to say, all those diverse areas of physics are based on 
the mathematics of two-by-two matrices.

Einstein introduced his special relativity nearly one hundred years ago.  
This theory of course revolutionized our understanding of space and time,   
and thereby introduced to physics a mathematical device called the Lorentz 
group.  Through its two-by-two representation, the Lorentz group is a 
very powerful instrument in theoretical physics.

\end{document}